# Nonlinear Coarse-graining Models for 3D Printed Multi-material Biomimetic Composites

Mauricio Cruz Saldívar[1*], Eugeni L. Doubrovski[2], Mohammad J. Mirzaali[1], Amir A. Zadpoor[1]

*1 Department of Biomechanical Engineering, Faculty of Mechanical, Maritime, and Materials Engineering, Delft University of Technology (TU Delft), Mekelweg 2, 2628 CD, Delft, The Netherlands*
*2 Faculty of Industrial Design Engineering (IDE), Delft University of Technology (TU Delft), Landbergstraat, 15, 2628 CE, Delft, The Netherlands*

---

[*] Corresponding author. e-mail: m.cruzsaldivar@tudelft.nl

**Abstract**

Bio-inspired composites are a great promise for mimicking the extraordinary and highly efficient properties of natural materials. Recent developments in voxel-by-voxel 3D printing have enabled extreme levels of control over the material deposition, yielding complex micro-architected materials. However, spatial complexity makes it a formidable challenge to find the optimal distribution of both hard and soft phases. To address this, a nonlinear coarse-graining approach is developed, where foam-based constitutive equations are used to predict the mechanics of biomimetic composites. The proposed approach is validated by comparing coarse-grained finite element predictions against full-field strain distributions measured using digital image correlation. To evaluate the degree of coarse-graining on model accuracy, pre-notched specimens decorated with a binarized version of a renowned painting were modeled. Subsequently, coarse-graining is used to predict the fracture behavior of bio-inspired composites incorporating complex designs, such as functional gradients and hierarchical organizations. Finally, as a showcase of the proposed approach, the inverse coarse-graining is combined with a theoretical model of bone tissue adaptation to optimize the microarchitecture of a 3D-printed femur. The predicted properties were in exceptionally good agreement with the corresponding experimental results. Therefore, the coarse-graining method allows the design of advanced architected materials with tunable and predictable properties.

## 1. Introduction

Extending and optimizing the design space of synthetic engineering materials remains an open problem. Overcoming this challenge is possible thanks to the additive manufacturing (AM, also known as 3D printing) of architected materials because the spectrum of achievable material properties can be dramatically extended by using single or multiple material phases in a wide range of geometrical configurations.[1–3] The emergence of these materials has, however, created new challenges because finding the proper design features at the smaller scales (*i.e.,* micro/nano) to create the desired and optimal combination of properties at a larger scale (*i.e.,* meso) is not trivial.[4]

In the recent past, researchers have used different approaches to design microarchitectures that lead to extraordinary macroscale properties.[5–9] One such approach is biomimetic design, where the essential characteristics of natural materials (*e.g.,* length-scale multi-hierarchical organization and functional grading) are used to rationally design a new class of engineered materials known as biomimetic architected materials (bio-AM). When designed optimally, such bio-AM exhibit some desirable combinations of ordinarily mutually exclusive properties (*e.g.,* high-strength and high-toughness).[10–15] Despite these advantages, the highly complex microarchitectures of bio-AM make their fabrication and analysis a formidable challenge.

Recent developments in the (multi-material) AM techniques have created new opportunities for fabricating such materials.[16–20] More specifically, voxel-based 3D printing has the potential to revolutionize architected materials.[21] This AM technique allows for a reliable "voxel-by-voxel" deposition of hard and soft material phases at the microscale, making it possible to mimic the spatial distribution of organic and mineral phases in natural composites [22,23] and provides an unlimited level of design freedom to create advanced bio-AM.[24–28] However, this approach can lead to exceptionally detailed and complex microarchitectures, complicating their nonlinear finite element analysis (FEA). The optimization of such designs is, therefore, only possible for small constructs because extensive simulations are typically needed to analyze every possible permutation of the hard and soft phases.[29–31]

A potentially useful approach to limit the computational cost associated with complex bitmap designs is to use coarse-graining.[4] This process has, for example, been used to represent monophasic porous materials (*e.g.,* bone, truss structures) by their grayscale equivalents, where the gray value represents the volume ratio of the solid phase. Although the classic implementations of this method do not necessarily result in manufacturable architectures, bitmap designs do not suffer from this issue. That is because the coarse-grained gray values of each representative volumetric element (RVE) directly translate to the volume fraction of the hard phase, $\rho$. The coarse-graining problem is, therefore, reduced to the problem of selecting the proper set of constitutive equations to serve as the gray value-property relationships of these composites while also adequately capturing their generalized nonlinear behavior.

Here, we developed a coarse-graining approach that allows for a highly efficient prediction of the nonlinear mechanical response of bio-AM. This methodology combines the design freedom of voxel-based AM with the simplicity of foam-based constitutive models to obtain the gray value-property estimation functions. This powerful approach circumvents the need for computationally expensive models and provides a highly efficient and cost-effective tool for the rational design of bio-AM. After performing only a few mechanical tests, we generated the data required to characterize the nonlinear constitutive equations and validated them via computational simulations. We then evaluated the effects of the coarse-graining degree by performing FEA on a highly-detailed portrait and gradually increasing the coarse-graining degree. Furthermore, we demonstrate applications of this approach by extending it to the analysis of the ductile fracture behavior of bioinspired materials.[25,32] Finally, we utilize the inverse coarse-graining operation to obtain the microarchitecture of a remodeled femoral bone.

## 2. Results and discussion

The results of a number of quasi-static tensile tests with different values of $\rho$ provided the data needed to characterize the gray value-property relationships for coarse-graining (**Figure 1**). Multi-material 3D printing based on the jetting of UV-curable polymers with voxel-level control over the type of the deposited material enables the generation of such tensile test specimens. To this end, stacks of binary images with randomly deposited white or black pixels represented the hard and soft phases, respectively (Figure 1A-B). Stacking these binary images made it possible to fabricate composites with reinforcement distributions in 3D (Figure 1C). This design freedom allowed including all the features required to 3D print the dogbone tensile tests specimens without additional post-processing (Figure 1D).

The behavior of the stress-strain curves resulting from the tensile tests ranged from hyperelastic (pure soft, $\rho = 0\%$) to elastoplastic (pure hard, $\rho = 100\%$) (**Figure 2**A). The ultimate strength achievable by these materials was between 3.1 and 62.1 MPa. The elastic modulus of the specimens varied between 0.9 and 2873 MPa (**Table S1** of the supplementary document) and increased nonlinearly in relation to $\rho$ (Figure 2B). These observations confirm the composite design range achievable by these materials, covering three orders of magnitude.[26] Therefore, to properly coarse-grain these materials, it is important to select a constitutive model that allows for retaining the characteristic curves and the magnitude of the changes that these composites present. Foam-based constitutive models for large deformations satisfy these conditions.[33]

They can represent both linear elastic and hyperelastic mechanical behaviors and can follow both hardening- or softening-like nonlinear regimes. The constitutive model used here defines the von Mises stress ($\sigma$) of a homogeneous material as:[33]

$$\sigma = A \frac{e^{\alpha\epsilon} - 1}{1 + e^{\beta\epsilon}} \tag{1}$$

In this equation, the constant $A$ has stress units, while the constants $\alpha$ and $\beta$ are dimensionless. Additionally, separating the contributions of the elastic ($\epsilon_{el}$) and plastic ($\epsilon_{pl}$) strains is possible with the expression:

$$\epsilon = \epsilon_{el} + \epsilon_{pl} = \frac{\sigma}{E} + \epsilon_{pl} \tag{2}$$

where the total von Mises strain ($\epsilon$) varies between zero and the ultimate strain ($\epsilon_{ult}$). Moreover, obtaining the elastic modulus ($E$) from equation 1 is possible after evaluating its derivative against the origin ($\frac{d\sigma(\epsilon=0)}{d\epsilon}$), resulting in the expression:

$$E = \frac{A\alpha}{2} \tag{3}$$

To generate the coarse-graining estimation functions, we characterized the parameters $A, \alpha, \beta$, and $\epsilon_{ult}$ in terms of $\rho$. To achieve this goal, we first obtained the equivalent parameters from our tensile test stress-strain results (Figure 2C). To more accurately represent the complex nature of the resulting curves, we generalized the parameters $A$ and $\alpha$ with two different power-law functions, with a transition region at $\rho = 40\%$ (**Table 1**). Similarly, we characterized the $\alpha/\beta$ ratio and ultimate strain $\epsilon_{ult}$ with exponential decay functions. These four simple parametrizations allowed us to accurately estimate the entire stress-strain behavior of any bio-AM for any $\rho$ value (Figure 2A, black curves).

To validate the coarse-graining equations, we developed FEA simulations of the tensile experiments (**Figure 3**). We changed the original resolution of the designs from 768×96×150 binary voxels to 128×16×25 greyscale RVEs. In this process, the gray value of each RVE represents the average $\rho$ value of the original voxels within it (**Figure S1** of the supplementary document). Then, we used the characterized coarse-graining functions to assign the respective mechanical properties to each RVE. As hexagonal elements were used to represent each RVE, the coarse-graining process reduced the element count by three orders of magnitude from 11.05×10^6 to 51.2×10^3, rendering the analyses computationally feasible.

Overall, the true principal logarithmic strain maps ($\epsilon_p$) resulting from digital image correlation (DIC) measurements were remarkably akin to those predicted by FEA (Figure 3A, **Figure S2**B of the supplementary document). Moreover, the behavior of the computational analyses adequately followed the transition from a highly elastic material with multiple cracks ($\rho = 10\% - 40\%$) to a more rigid material that presents necking ($\rho = 60\% - 100\%$). This high precision across the entire design space was also present in the comparisons between the properties estimated by the FEA and those obtained in the experiments (Figure 3B). In those, the elastic modulus and ultimate strength properties were very highly correlated ($R^2 > 98\%$), with slight underpredictions of the elastic modulus for the intermediate values of $\rho$. Additionally, the stress-strain curves of the computational models followed the hardening- or softening-like behaviors of their respective tests (Figure S2C of the supplementary document). However, FEA overpredicted the $\epsilon_{ult}$ and toughness ($U_d$) for the tests with $\rho < 50\%$ and underpredicted them for $\rho > 50\%$. These differences are likely caused by the processes involved in the DIC measurements and can be reduced by performing the DIC recordings at the micro-scale, leading to a higher precision than the macro-level recordings of this study. Regardless of these differences, the $R^2$ values characterizing the agreement between our numerical data and the experiments approached unity. The qualitative, characteristic features of the curves and the ductile failure mechanisms of the bitmap composites were also retained in the coarse-grained simulations. These observations confirm the accuracy and utility of the presented coarse-graining approach for capturing the nonlinear behavior of voxel-based 3D printed composites.

To further validate the application of our coarse-graining process, we evaluated how the degree of coarse-graining affects the overall FEA performance in a process similar to a mesh convergence study. To consider a structure with more refined local features, we used a part of the famous portrait *Girl with a pearl earring* by Johannes Vermeer (Delft, c. 1665) (**Figure 4**) to generate a 3D bitmap composite. We simulated the mechanical response of the portrait-decorated design in two different FEA analyses, each with a different initial binary resolution. The first set of the simulations comprised a set of linear elastic simulations, where the original binary 3D image was represented by 288×288 voxels in the plane and six voxels in depth and was coarse-grained to different degrees (Figure 4A). The second set of simulations was performed until the yield point and had an original resolution of 126×126 voxels in the plane and 4 voxels in depth (Figure 4B). We then compared the outcomes of these simulations and the required computational resources to better understand the effects of the coarse-graining degree.

The original binary images and the coarse-grained specimens shared the same regions of strain concentration (Figure 4A-B). The strains concentrated mainly at the forehead, chin, and cheek regions, which are the regions with abrupt transitions in the material type. However, the simulations at higher resolutions showed more heterogeneous strain patterns and had higher local strain values than the coarse-grained ones. Although these effects must be acknowledged and considered when coarse-graining designs with fine features (*e.g.*, under the presence of a crack front or when abrupt material transitions are present), the local strain distributions did not substantially affect the overall response of the architected materials (Figure 4C). When comparing the overall mechanical properties calculated using the different models, the prediction error for the elastic modulus was always <13%. At the same time, the yield strength did not deviate more than 17% from the estimates obtained using the original model. Although these deviations may be critical depending on the application type, one must consider the efficiency gained from the process. The required computational time dropped (Figure 4D) by three orders of magnitude such that the most coarse-grained models required only 0.45% and 0.3% of the original time to compute the elastic modulus and yield stress, respectively. Such a huge drop in the required computational cost can change the fate of an intended simulation from "practically impossible" to "routinely possible", making it possible to take design steps (*e.g.*, optimization iterations) that are otherwise infeasible. The price paid in terms of the lost accuracy should, however, be carefully considered for each specific application.

To harness the advantages of the proposed coarse-graining approach for simulating large and complex bio-AM structures, we simulated the ductile mode I crack propagation of designs from two of our previous studies [25,32]. These were three different bioinspired designs and three crack designs with randomly distributed particles ($\rho = [100\%, 75\%, 50\%]$) (**Figure 5**A, **Figure S3** of the supplementary document). The bioinspired designs included a functional gradient (FG) and two hierarchically organized (*i.e.,* with one and two hierarchy levels) brick and mortar designs with embedded FGs (BMG-1L and BMG-2L). Initially, these designs had a resolution of $1728 \times 864 \times 111$ voxels, which we coarse-grained to $864 \times 432 \times 1$, thereby making 2D plane stress computational analyses feasible. These models allowed us to study the differences between the strain distributions and the macroscopic properties (*i.e.,* elastic modulus and ultimate strength) calculated using the coarse-grained FE models and the effective mechanical properties and full-field strain distributions measured experimentally for the original bio-AM designs.

The strain distributions of the coarse-grained models were similar to those measured using DIC (Figure 5B-E, Figure S3 of the supplementary document). In the case of the particle-distributed designs ($\rho = 50\%, 75\%, 100\%$), both the DIC and FEA strains presented the butterfly-shaped crack tip regions usually observed in pre-notched mode I specimens. In contrast, the strains of the FG design followed a horizontal pattern ahead of the crack tip, concentrating in the region with the softer material. Furthermore, both the BMG-1L and BMG-2L designs presented the small horizontal and vertical strain concentration lines that are characteristics of the brick and mortar designs [25] and combined the previously described butterfly and horizontal strain distribution patterns. These observations further confirm that the proposed coarse-graining method can successfully capture the original features of complex bio-AM designs. Similar to the strain distributions, the effective behavior predicted with FEA followed the same trends observed in the experiments (Figure 5F). The predicted values were, however, somewhat lower than the measured ones. In the case of the elastic modulus, the underestimations were mostly present when the concentration of the hard phase around the crack tip was low (*e.g.,* the designs with $\rho = 50\%$, FG). Since the ductile fracture criterion used here is based on the ultimate strain (and not ultimate stress), an underestimation of the elastic modulus also led to an underestimation of the strength. Future studies should evaluate the use of other types of ductile fracture criteria. Despite these limitations, the ranking and the rough values of the estimated quantities matched the experimental results. These agreements confirm that the coarse-graining approach presented here can capture the effects of complex hierarchical design features reasonably well, making it suitable for the design of micro-architected bio-AM composites.

The functionality and precision offered by the coarse-graining approach are not only limited to the simplification of large-scale and complex bitmap structures. Given the accuracy of the gray value-property relationships, we hypothesized that it should be possible to use their inverse operation to back-calculate the bio-AM microarchitectures of grayscale structures without changing the expected mechanical response. Therefore, we extended our analyses by using this inverse coarse-graining operation to discretize, manufacture, and test the voxel structures obtained from a bone remodeling process.[34] For that purpose, we considered a 2D proximal femur that initially contained a homogeneous distribution of 50% hard material (**Figure 6**A). We used FEA to numerically simulate the femur under force-controlled conditions. For optimization, we updated every element's $\rho_i$ value by comparing its normalized strain energy density (normalized with respect to the maximum value of the same quantity, Figure 6B) ($S_i$) with the mean normalized strain energy density of the structure ($S_{avg}$) at the end of each iteration. After the convergence of the remodeling process (*i.e.,* after 42 iterations, Figure 6C), we 3D printed and tested the first and last iterations. To perform the inverse coarse-graining operation, we transformed every greyscale RVE into a random group of 3×3×3 binary voxels (*i.e.,* a change in the resolution from 206×118×1 to 618×354×3 voxels), where the number of the hard voxels depended on the $\rho$ value of each RVE. We repeated this voxel assignment process 144 times to generate all the 432 different bitmap images required to manufacture these femora. Finally, we 3D printed these designs and tested them under uniaxial compression until failure.

The force-displacement plots proved that the remodeling process enhanced the total energy capacity of the designed bone (Figure 6D). This process estimated a maximum force improvement of 146% (from 418.43 N to 612.85 N) while reducing the entire stiffness of the system from 485.09 Nmm$^{-1}$ to 205.67 Nmm$^{-1}$ (42.4%). Similar improvements were also present in the experimental results, where the maximum force increased by 139.2% (from 516.67 N to 719.27 N), accompanied by a 42.76% reduction in the stiffness (from 518.25 Nmm$^{-1}$ to 221.61 Nmm$^{-1}$). These results indicate that FEA underestimated the experimental outcome. The FE-predicted stiffness values were between 92.8% and 93.6% of the corresponding experimental values and the FE-predicted maximum force ranging between 81.0% and 85.2% of the corresponding experimental values. Despite these differences, the percentages of the property changes between the last and first iterations were very close for the FEA predictions and experimental measurements (*i.e.,* 146% *vs.* 139.2% for the maximum force and 42.4% *vs.* 42.76% for the stiffness). Moreover, the patterns of the changes in the strain distributions, which improved the overall properties of the designs, were virtually indistinguishable between the FEA results and the DIC measurements (Figure 6E). After concluding the optimization process, the designed femur exhibited high strain concentrations across the head, neck, and greater trochanter regions, suggesting an appropriate and efficient internal load distribution.[35] In contrast, the initial iteration only showed strain concentrations in the diaphyseal regions. These results highlight the potential of the proposed inverse coarse-graining approach as a tool to design, optimize, and discretize bio-AM, enabling the development of highly efficient and miniaturizable (medical) devices with advanced functionalities.

## 3. Conclusion

In summary, we showed that foam-based constitutive equations for large deformations effectively and accurately model the nonlinear mechanical behavior of bio-AM. This simple and elegant system of functions combined with voxel-by-voxel 3D printing techniques provides the user with a methodology to coarse-grain large systems of binary images and represent them at smaller greyscaled resolutions to perform nonlinear quasi-static numerical analyses and optimization processes. The coarse-graining and numerical modeling of the detailed bitmap designs substantially reduced the simulation time by three orders of magnitude while only causing reasonably small deviations from the true mechanical response. Thanks to this huge reduction in the simulation time, we could perform an analysis of large and complex bio-AM to study their local strain distributions and estimate their performance. Furthermore, the proposed coarse-graining approach enabled us to mimic the internal material distribution of a femoral bone that results from a remodeling algorithm, yielding a complex local strain distribution that enhanced the overall nonlinear response of the structure. The inverse coarse-graining process enabled discretizing these greyscaled structures into manufacturable equivalents while maintaining the outstanding mechanical response estimated by their computational counterparts. The proposed methodology, therefore, creates new opportunities for designing advanced materials given their desired properties in a reliable and straightforward manner. Many high-tech industries, such as soft robotics, aerospace, orthopedics, and tissue engineering, may benefit from such a simulation approach that could enable them to incorporate more voxel-based approaches into the design of their architected materials.

## 4. Materials and methods

*3D printing:* We used a multi-material jetting 3D printer (ObjetJ735 Connex3, Stratasys® Ltd., USA) with UV-curable photopolymers to fabricate the specimens. The deposition of each polymer type (*e.g.*, hard *vs.* soft) could be controlled at the level of individual voxels. Stacks of binary images were, therefore, required as input. Each material phase required its own stack of images, wherein the white bits (*i.e.*, 1) represent the location where the 3D printer will deposit that material particle. The resolution of the machine was 600×300 dpi in layers of 27 μm, which yields voxels with the native dimensions of 42×84×27 μm$^3$. We selected the photopolymers VeroCyan$^{TM}$ (RGD841, Stratasys® Ltd., USA) and Agilus30$^{TM}$ Clear (FLX935, Stratasys® Ltd., USA) as the hard and soft phases, respectively. We processed the prints with GrabCAD Print (Stratasys® Ltd., USA), but the designs and stacks of the binary images were prepared in MATLAB R2018b (Mathworks, USA). We aligned the designs and bitmap images along the same axis as the printing direction, thereby minimizing the inclusion of printing-related anisotropy in our specimens.

*Design and testing of the tensile test specimens:* We defined the tensile test specimens with seven values of the volume fraction of the hard phase (*i.e.*, $\rho = 0, 10, 20, 40, 60, 80, 100$ %, Figure 1A). We then randomly distributed a large enough number of voxels to discretize the internal architecture of these designs. We projected these designs onto the standard tensile test specimen shape (type IV) described in ASTM D638-14 (thickness = 4 mm) (Figure 1D).[36] A particular approach was followed for the pure soft (*i.e.*, $\rho = 0\%$) specimens since we only applied the design at at the centermost 8 mm of the gauge length while connecting the edges with a monotonic gradient of $\rho$ (Figure S2A of the supplementary document). We assigned the hard phase to the gripping part of each specimen. Three specimens were then 3D printed from each of the abovementioned designs. We used a mechanical testing machine (LLOYD instrument LR5K, load cell = 5 kN) to deform the specimens at a rate of 2 mm min$^{-1}$ until failure. The measured time ($t$) and force ($f$) signals were recorded at a frequency of 100 Hz. To improve the accuracy of our measurements, we recorded the local displacement fields of every test by using a DIC system (Q-400 2x 12 MPixel, LIMESS GmbH, Krefeld, Germany) at a frequency of 1 Hz. Towards that aim, we applied a black dot speckle pattern over a white paint background on every specimen prior to testing. Then, we used a commercial program (Instra 4D v4.6, Danted Dynamics A/S, Skovunde, Denmark) to calculate the full-field strain maps (equivalent von Mises true strains and first principal true strains) of each tested specimen. To extract the numerical data and to characterize the behavior of the composites, we used a point digital extensometer at the failure location of each specimen to extract the corresponding vectors of true von Mises strains ($\epsilon$). Similarly, to obtain data for the FEA validations of these tests, we used a line extensometer that spanned the entire gauge length of the specimen and extracted the respective true strain vectors. Then, we used MATLAB R2018b to calculate the true stresses ($\sigma = f A_o^{-1} \exp(\epsilon)$, $A_o = 32.512$ mm$^2$) and elastic modulus, $E$, of the specimens. To determine the elastic modulus, a line was fitted between 0% and 30% of the maximum recorded stress. The ultimate tensile strength, $\sigma_{max}$, was determined as the maximum recorded stress while the strain at failure, $\epsilon_{ult}$, was defined as the last recorded strain. Finally, the strain energy density, $U_d$, was calculated as the area under the stress-strain curve of each specimen.

*Characterization of Foam based constitutive models:* The original foam-based constitutive model for large deformations chosen for this study has the form: [33]

$$\sigma = A \frac{e^{\alpha\epsilon} - 1}{B + e^{\beta\epsilon}} \qquad (A)$$

In this function, *A* has the unit of stress and represents the asymptotic value of the stress that is reached when $\alpha$ and $\beta$ are equal. $\alpha$ and $\beta$ are the dimensionless constants that influence how fast the stress ($\sigma$) will reach the asymptotic value *A*. Their ratio ($\alpha/\beta$) determines whether the curve is hardening-like or softening-like. *B* is a dimensionless parameter that generates a toe region in the curve. We defined this latter parameter as unity (*B=1*) for simplicity, yielding the function presented in equation 1. This equation represents the entire history of the von Mises stress for a composite and is directly related to the respective von Mises strains ($\epsilon$) until the ultimate strain ($\epsilon_{ult}$) is reached. To use this equation in FEA, one needs to separate the plastic strain ($\epsilon_{pl}$) from the elastic ($\epsilon_{el}$) strain. Assuming that the elastic regime is linear elastic for all cases, it is possible to derive the following equations:

$$\epsilon = \epsilon_{el} + \epsilon_{pl} \qquad\qquad (B)$$

$$\epsilon_{el} = \sigma/E \qquad\qquad (C)$$

Equation 2 is easily obtained after combining these two expressions. Furthermore, since the elastic modulus ($E$) is the initial slope of the stress-strain curve, it can be obtained by differentiating equation 1 with respect to strain and evaluating the resulting function at the origin:

$$\frac{d\sigma}{d\epsilon} = A\frac{\alpha e^{\alpha\epsilon}\left(1 + e^{\beta\epsilon}\right) - \beta e^{\beta\epsilon}(e^{\alpha\epsilon} - 1)}{(1 + e^{\beta\epsilon})^2} \qquad\qquad (D)$$

$$E = \frac{d\sigma(\epsilon = 0)}{d\epsilon} = A\frac{\alpha(1 + 1) - \beta(1 - 1)}{(1 + 1)^2} = \frac{A\alpha}{2} \qquad\qquad (E)$$

To generate the input data to characterize the parameters of the constitutive model, we used nonlinear least squares to fit equation 1 to each of the stress-strain curves of the tensile test specimens. Then, we used the resulting values of the parameters $A$, $\alpha$, $\beta$, and $\epsilon_{ult}$ and their respective $\rho$ values to generate the desired gray value-property functions. We also used nonlinear least squares to fit the functions expressed in Table 1 to these parameters.

*FEA of tensile test experiments:* We used a commercially available nonlinear solver (*i.e.,* Abaqus Standard v.6.14, Dassault Systèmes Simulia, France) to create quasi-static finite element models of the gauge length of the tensile test specimens (the narrow region). We used hexagonal hybrid elements with reduced integration (C3D8RH) for discretizing each RVE of the coarse-grained designs. We used the gray value-property relationships obtained in the previous step and the $\rho$ value of each RVE to calculate their respective mechanical properties. Additionally, we assigned element deletion via ductile fracture, where the ultimate plastic displacement of every element was obtained from the ultimate strain function and equation 2. The specimens were loaded by prescribed uniaxial displacement as the boundary condition at the one end of the specimens and applying fixed displacement (encastre) boundary conditions on the cross-section of the other end. The outputs of the computational models were the forces and displacements of the structure, which we converted into true stresses and strains. The predicted mechanical properties corresponding to the measured values were then calculated using a procedure similar to the one used in the experimental data analysis. For validation, we made one-to-one comparisons between the FEA results and the available experimental data and calculated the corresponding coefficients of determination to evaluate the accuracy of our models.

*Design and FEA of portrait specimens:* We discretized a grayscale copy of a part of the renowned painting *Girl with a pearl earring* into two different 3D binary images with different resolutions (*i.e.,* 288×288×6 and 126×126×4) (Figure 4). To binarize the image, we first reduced its resolution to the mentioned 2D resolutions and assigned the number of hard voxels randomly depending on the local $\rho$ values. We then applied this process to each layer of every design (*i.e.,* 6 or 4 times) and stacked the results to generate the 3D binary images. To generate the FEA meshes, we coarse-grained these images at different resolutions. For the simulations focused on determining the elastic modulus (*i.e.,* the design with a native resolution of 288×288×6 voxels), we coarse-grained the portrait to the following resolutions: 144×144×6, 96×96×6, and 48×48×6 (greyscale RVEs in all cases). To calculate the yield stress (*i.e.,* the design with a native resolution of 126×126×4 voxels), coarse-graining resulted in the following resolutions: 72×72×4, and 36×36×4 (greyscale RVEs). We represented each RVE with four hexagonal hybrid elements with reduced integration (C3D8RH). These discretizations resulted in meshes with $1.99×10^6$, $4.97×10^5$, $2.21×10^5$, and $5.52×10^4$ elements for the simulations aimed at determining the elastic modulus and $2.54×10^5$, $6.35×10^4$, and $1.58×10^4$ for those performed to calculate the yield stress. The mechanical properties assigned to each RVE were determined using the above-derived gray value-property relationships. We then used these models for performing some simulations that resembled quasi-static tensile tests by applying a constant vertical displacement on the top end of the specimens while constraining all the displacements at the bottom end of the specimens. The simulations results were subsequently used to calculate the elastic modulus and maximum strength for all the analyses. The computational time required to run each simulation was also measured.

*Design, testing, and FEA of bioinspired pre-notched specimens:* We used six different bioinspired pre-notched specimen designs that had a total of 1728×864×111 binary voxels, which is equivalent to the following dimensions: $73.152×73.152×2.997$ mm$^3$. These structures included a single-edged crack at the middle of their length that spanned 20% of their width. We used the hard phase to define the gripping part of each specimen according to their original studies.[25,26] We then 3D printed and tested these specimens using the same equipment and methodology as described above for the tensile test specimens. For post-processing, we created virtual strain gauges on the DIC results that encompassed the entire region of each design to extract the engineering strain vectors of every specimen. We used the force vectors obtained from the testing machine to calculate the respective engineering stress vectors ($\sigma = f A_o^{-1}, A_o = 219.23$ mm$^2$) and to calculate the elastic modulus and maximum strength for each specimen. After post-processing the experimental results, we modeled these tests using finite element models. To generate the geometries used in the models, we coarse-grained the specimens from 1728×864×111voxels to 864×432 RVEs and used the previously obtained gray value-property relationships to determine the mechanical properties that were subsequently assigned to the RVEs. We used quadrilateral plane stress elements with reduced integration (CPS4R) for meshing and enabled element deletion via ductile failure, where the ultimate plastic displacement of every element was obtained from the ultimate strain function and equation 2. As for the boundary conditions, we prescribed uniaxial displacement until separation at the top end of the specimens while constraining all the movements of the bottom surface. We made one-to-one comparisons between the predicted and measured mechanical properties while also comparing their corresponding stress-strain curves. We also calculated the coefficients of determination between the modeling and experimental results to assess the accuracy of the finite element predictions.

*Design, remodeling, testing, and FEA validation of a femoral bone:* The original femoral bone design consisted of a long femur bone STL obtained from a free CAD website,[37] which we converted into a 3D image using a program provided by Adam A.[38] The final 2D image was a slice cut on the frontal plane of the proximal section of the femur. The resolution of the image was 206×118 voxels. For the remodeling process, the FEA simulations were performed using a 2D mesh of quadrilateral plane strain elements with reduced integration (CPE4R) and were loaded according to Figure 6A. After every remodeling iteration, we updated the $\rho_i$ value of every element (*i.e.,* subindex $i$) depending on their normalized strain energy density ($S_i$), inspired by the homeostasis-based approaches for modeling the bone tissue adaptation process.[39] We defined the normalized strain energy density with the following expression:

$$S_i = \frac{U_i}{U_{max}(\rho_i)} \tag{F}$$

In this expression, $U_i$ is the strain energy density of each element while $U_{max}(\rho_i)$ is the maximum value of the strain energy density that the material can withstand (Figure 6B) and is obtained by integrating the stress-strain curve:

$$U_{max}(\rho_i) = \int_0^{\epsilon_{ult}} A\frac{e^{\alpha\epsilon} - 1}{1 + e^{\beta\epsilon}} d\epsilon \tag{G}$$

For every iteration, we updated the rate of the change in the volume fraction of the hard phase ($\dot{\rho}_i$) corresponding to each element using the following relationships:

$$S_{avg} = \sum_{i=1}^{N} \frac{S_i}{N} \tag{H}$$

$$\dot{\rho}_i = \left(\frac{S_i}{S_{avg}} - 1\right)\dot{\rho}_s \qquad \text{(I)}$$

where $S_{avg}$ is the average normalized strain energy density across the entire bone and $\dot{\rho}_s = 0.05$ is the slope of the rate of change function. Then, we incorporated a "lazy zone" in which no change in the hard volume fraction occurs and truncated the upper and lower boundaries of $\dot{\rho}_i$ possible in each iteration. These processes involve adding the following conditional expression:

$$\dot{\rho}_i = \begin{cases} 0 & |\dot{\rho}_i| < \dot{\rho}_{lazy} \\ \dot{\rho}_i & \dot{\rho}_{lazy} \leq |\dot{\rho}_i| \leq \dot{\rho}_{max} \\ \pm\dot{\rho}_{max} & |\dot{\rho}_i| > \dot{\rho}_{max} \end{cases} \qquad \text{(J)}$$

$$\rho_i = \rho_i + \dot{\rho}_i \qquad \text{(K)}$$

where $\dot{\rho}_{lazy} = 0.005$ is the limit value of $\dot{\rho}$ where no change of $\rho$ occurs and $\dot{\rho}_{max} = 0.1$ is the maximum (or minimum) value of $\dot{\rho}$ enabled for each iteration. Finally, we updated the $\rho_i$ values corresponding to each element using the following relationship:

$$\rho_i = \rho_i + \dot{\rho}_i \qquad \text{(L)}$$

We continued this iterative remodeling process until the total change of hard material volume fraction ($\sum_{i=1}^{N} |\dot{\rho}_i|$) dropped to 5% when compared to the corresponding value of the initial iteration (Figure 6C). Consequently, a total of 42 remodeling iterations were required for convergence. Once the bone remodeling concluded, we performed the above-described inverse coarse-graining operation to obtain the 3D printable geometries of the specimens corresponding to the first and last iterations. We tested these designs under quasi-static compression under the same conditions and with the same equipment as the tensile tests and extracted the resulting forces and displacements. For a final validation of the coarse-graining equations, we performed additional FEA of these tests under the same compression loading conditions. For this purpose, we coarse-grained back the 618×354×432 voxels of these designs into an equivalent of 206×118×4 RVEs and discretized the resulting geometries using the C3D8RH elements.

**Competing interests**

The authors declare no competing interests.

**Acknowledgments**

powerful platform for experimentation, discovery, and innovation; for more information contact: academic.research@stratasys.com.

**Supplementary material**

Please see the supplementary file of this project.

**Data availability**

The raw and processed data used in this study will be made available upon request from the authors.

**Table 1.** The foam-based constitutive equation parameters used to coarse-grain bio-AM in terms of the local value of the volume fraction of the hard phase ($\rho$).

| Parameter | $\rho \leq 40\%$ | $\rho > 40\%$ |
|---|---|---|
| $A$ [MPa] | $A = 14.951\rho^{0.882} + 0.695$ | $A = 67.980\rho^{1.272} - 14.015$ |
| $\alpha$ [-] | $\alpha = 642.599\rho^{3.36} + 2.035$ | $\alpha = 4778.62\rho^{0.0157} - 4678.65$ |
| $\alpha/\beta$ [-] | $\alpha/\beta = 0.892e^{-7.52\rho} + 1.0037$ | |
| $\epsilon_{ult}$ [mm mm$^{-1}$] | $\epsilon_{ult} = 0.698e^{-9.218\rho} + 0.452$ | |

**Figure captions**

**Figure 1.** A) The representative binary images used to design and fabricate the tensile test specimens. These were 3D printed with a multi-material 3D printer and contained different values of the volume fraction of the hard phase ($\rho$). B) The stress-strain curves of monolithically soft and hard specimens. C) A representative volume element of a 3D printed specimen designed by voxel-based 3D printing. These composites contained randomly distributed hard and soft voxels. D) A standard tensile test specimen implementing the random design within its gauge length (out-of-plane thickness = 4 mm).

**Figure 2.** A) The average stress-strain curves of the tensile tests. The shaded area around the mean curves represents ±SD. The different line and color gradients refer to the specimens with different $\rho$ values. The dotted black lines show the predictions of the coarse-graining functions. B) A comparison between the experimental values of the elastic modulus and the corresponding estimations obtained using the coarse-graining equations. C) The average values of the experimental (Exp.) *vs.* fitted parameters that define the constitutive model represented in equation 1 (*i.e., A, α*, and $\alpha/\beta$) and ultimate strain before failure ($\epsilon_{ult}$) used to coarse-grain the bitmap composites.

**Figure 3.** A) The experimental (Exp.) *vs.* FEA results for the elastic modulus ($E$), ultimate tensile strength ($\sigma_{max}$), ultimate strain ($\epsilon_{ult}$), and strain energy density (U$_d$) obtained from the quasi-static tensile tests for all the specimens, together with the corresponding coefficients of determination ($R^2$). B) The distribution of the first principal true strain ($\epsilon_p$) measured experimentally using DIC and calculated with coarse-grained FEA of the same specimens. All the images refer to the last time point before critical separation.

**Figure 4.** The effects of the degree of coarse-graining on the strain distributions and the effective mechanical properties of the specimens. These simulations were performed on a binarized copy of a part of the renowned painting *Girl with a pearl earring* (Johannes Vermeer, Delft, c. 1665), where coarse-graining was performed to different degrees. A) The distribution of the principal true strain calculated using linear elastic finite element models. The original model had a resolution of 288×288 binary bits and was coarse-grained from $1.99 \times 10^6$ (*i.*) to $5.52 \times 10^4$ (*ii.*) hexagonal FEA elements. B) The strain distribution predicted by the finite elements that simulated the loading of the specimens until yielding occurred at the crack tip. The original model had a resolution of 126×126 binary bits and was coarse-grained from $2.54 \times 10^5$ (*iii.*) to $1.58 \times 10^4$ (*iv.*) hexagonal FEA elements. C) The elastic modulus and yield

strength of calculated using the finite element models. D) The computational time *vs.* the number of linear hexagonal elements used in these simulations.

**Figure 5**. A) As a case study, we selected several bio-AM microstructures with complex features for modeling. These designs were compared with a pre-notched specimen with a $\rho$ of 100%. A functionally graded (FG) specimen, as well as two FG designs embedded with brick and mortar designs with one (BMG-1L) and two (BMG-2L) levels of hierarchical organization were also included. The predicted values of the principal strain are compared with the DIC results for the B) $\rho = 100\%$, C) FG, D) BMG-1L, and E) BMG-2L specimens. F) A comparison between the measured (Exp) and predicted behaviors of the different designs.

**Figure 6.** The optimization of the femoral microarchitecture using a theoretical model of the bone tissue adaptation process and implementing an inverse version of the proposed coarse-graining approach. A) The initial FEA remodeling setup and boundary conditions. B) The maximum value of the strain energy density ($U_{max}$) that an element can withstand as a function of $\rho$. C) The convergence curve of the remodeling process, where the total rate of change in the volume fraction of the hard phase ($\Sigma |\dot{\rho}_i|$) was calculated for every iteration. D) The force-displacement curves corresponding to the compression tests performed on the starting and final microarchitectures resulting from the optimization process (both measured and predicted values). E) A comparison between the designs and FEA strain field distributions of the grayscale models and those from the discretized bitmap structures obtained from the inverse-coarse graining process.

**Figure 1**

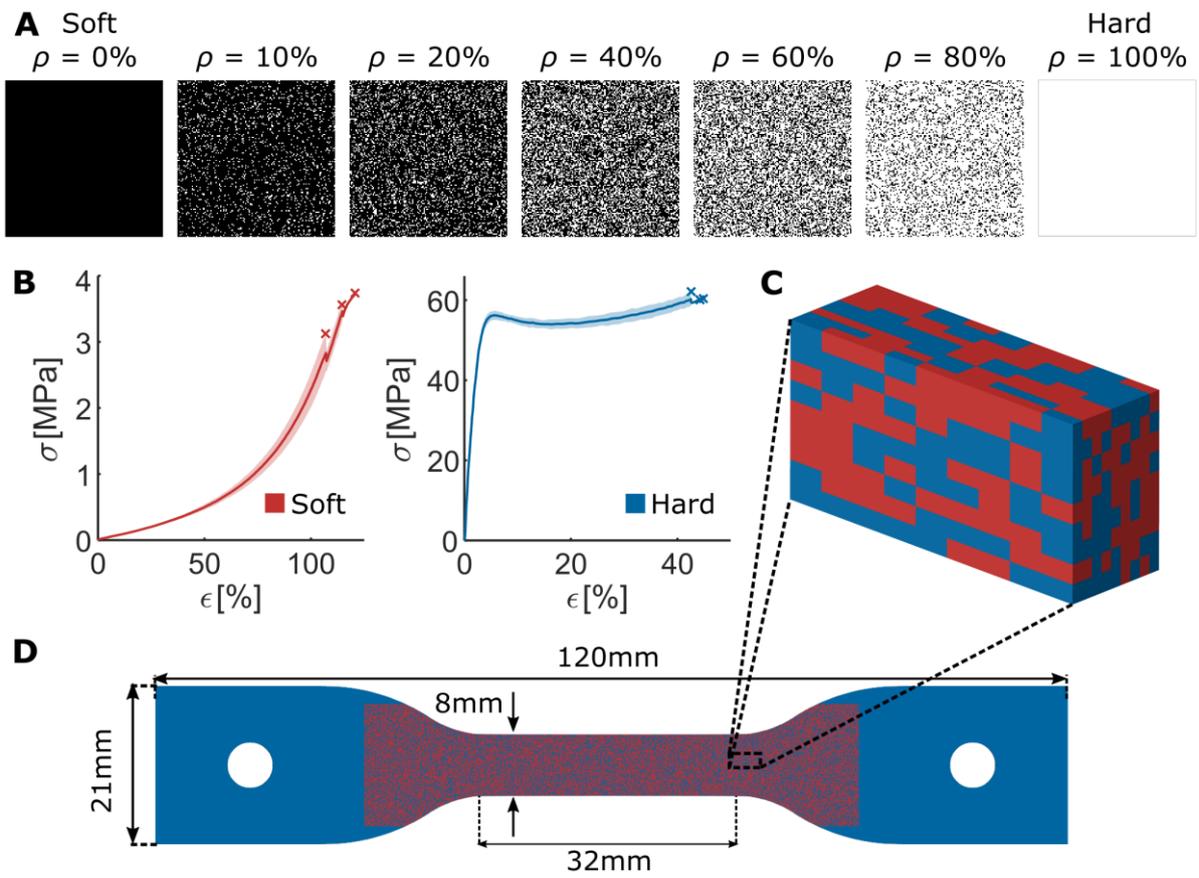

A  Soft
$\rho = 0\%$  $\rho = 10\%$  $\rho = 20\%$  $\rho = 40\%$  $\rho = 60\%$  $\rho = 80\%$  Hard
$\rho = 100\%$

B

C

D  120mm

21mm  8mm

32mm

**Figure 2**

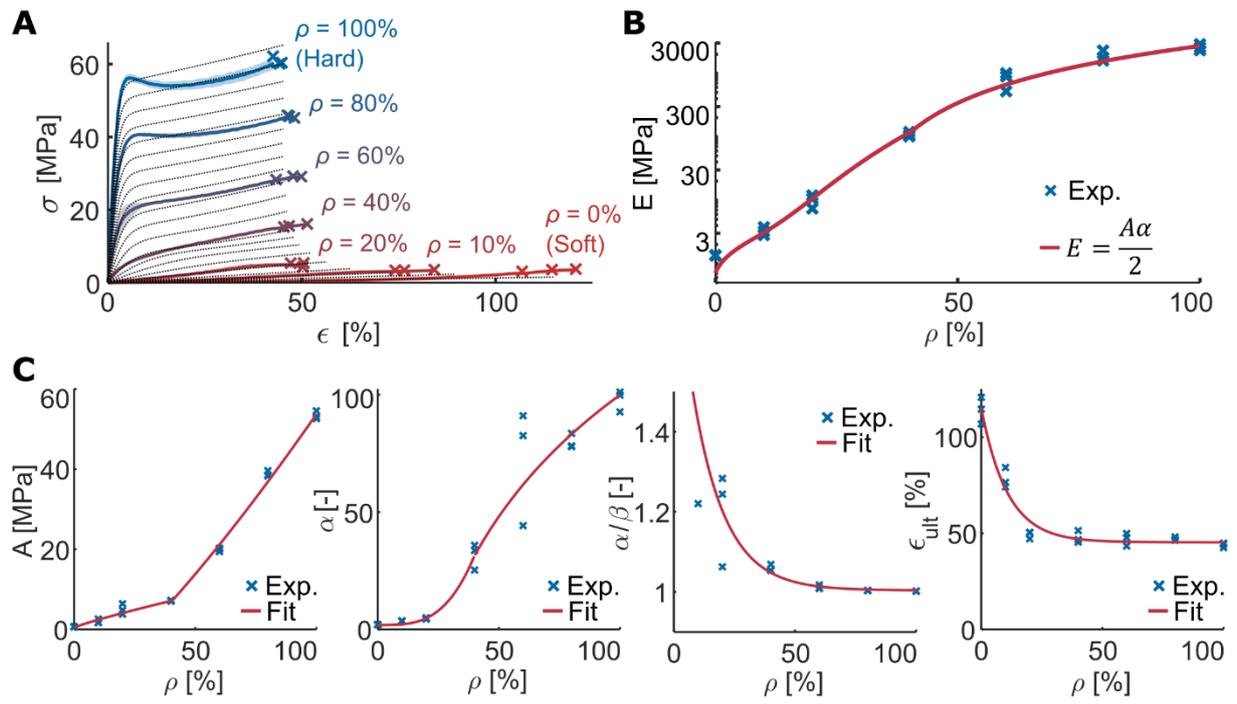

**Figure 3**

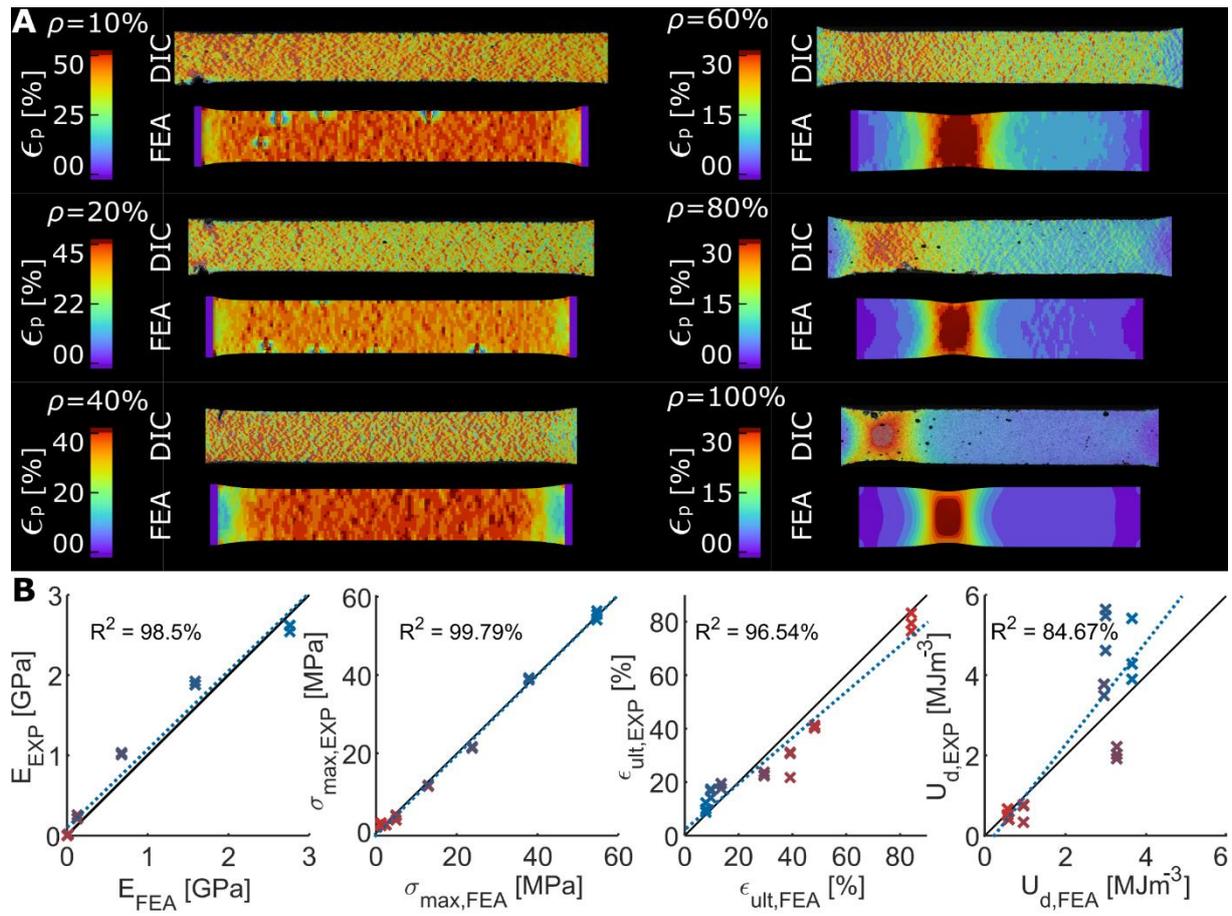

**Figure 4**

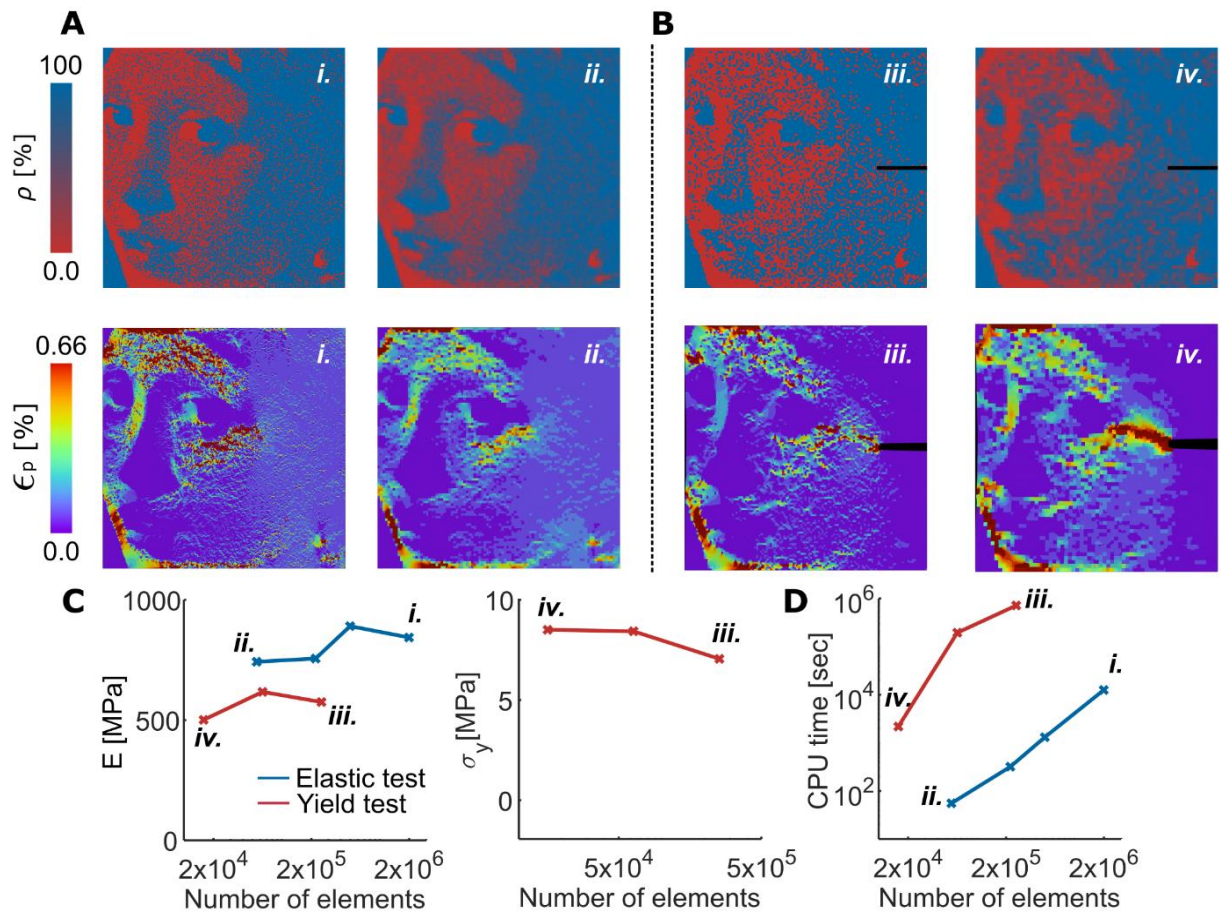

**Figure 5**

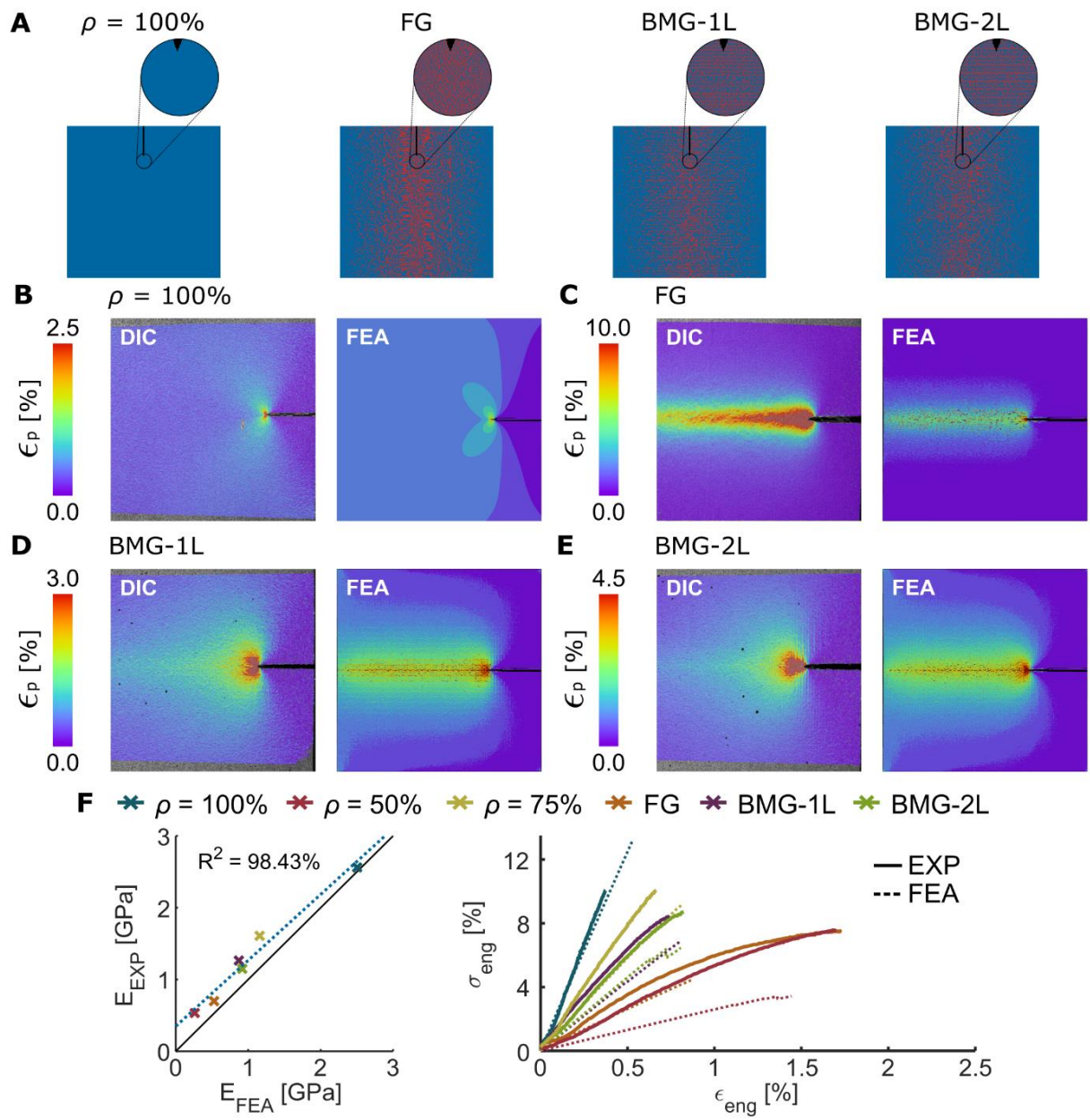

**Figure 6**

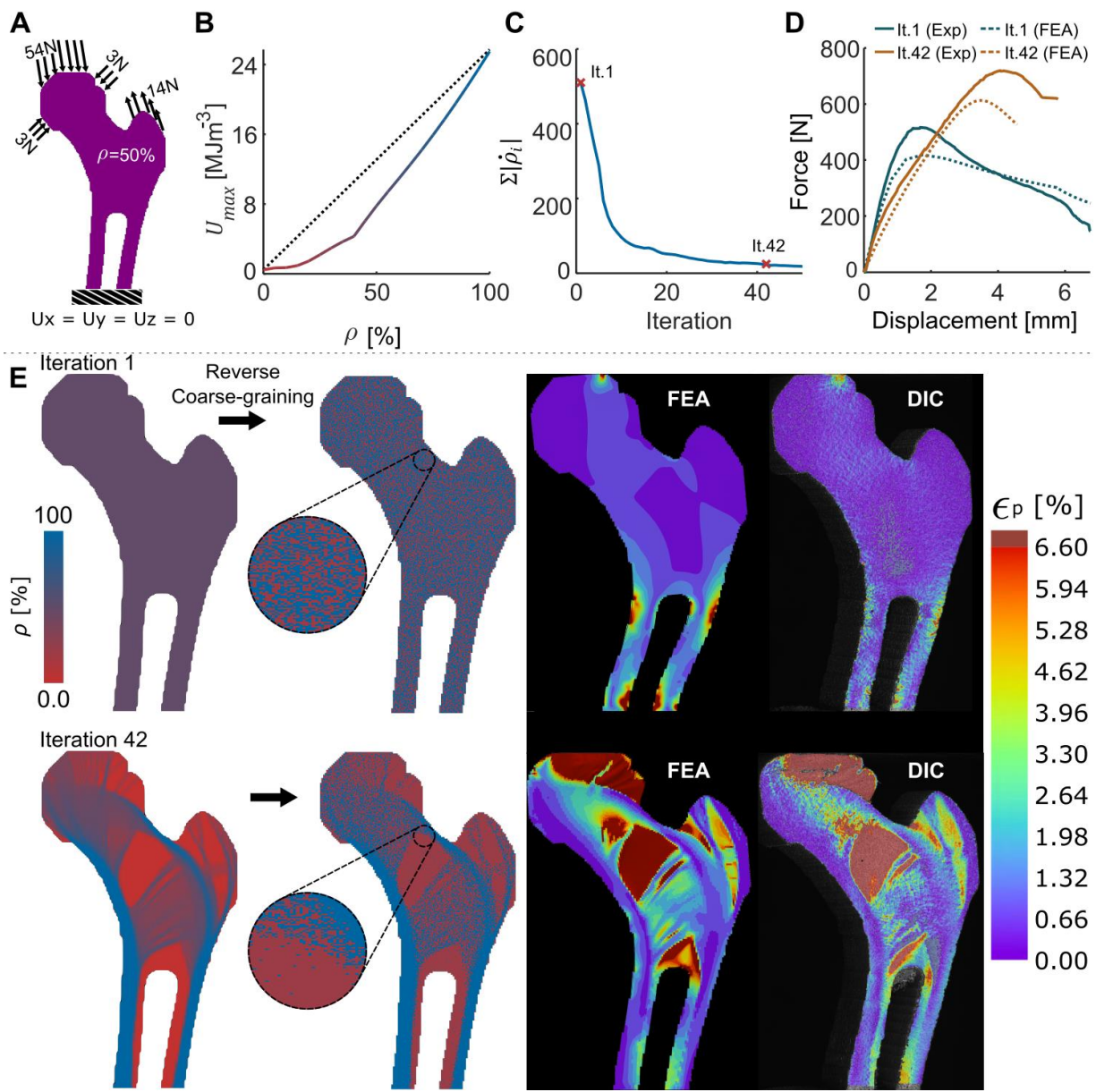